\documentclass[a4paper]{article}
\usepackage{graphicx}
\usepackage{amsmath}
\usepackage{amsfonts}
\usepackage{amssymb}

\begin{document}

\author{V. A. Ivanov, M. van den Broek and F. M. Peeters\\Departement Natuurkunde, Universiteit Antwerpen (UIA),\\Universiteitsplein 1, B-2610 Antwerpen, Belgium}
\title{Strongly interacting $\sigma$-electrons and MgB$_{2}$ superconductivity}
\date{25th April 2001}
\maketitle
\begin{abstract}
MgB$\left(  p^{n_{\sigma}}p^{n_{\pi}}\right)  _{2}$ is classified as a system
with strongly interacting $\sigma$-electrons and non-correlated $\pi
$-electrons of boron ions. The kinematic and Coulomb interaction $V$ between
the orbitally degenerated $\sigma$-electrons provide the superconducting state
with an anisotropic gap of $s$*-wave symmetry. The critical temperature
$T_{c}$ has a non-monotonic dependence on the distance $r$ between the centers
of gravity of $\sigma$- and $\pi$-bands. MgB$_{2}$ corresponds to $r=0.085$ eV
\ and $V=0.45$ eV in our model with flat bands. The derived superconducting
density of electronic states is in good agreement with available experimental
and theoretical data. The possibilities for increasing $T_{c}$ are
discussed.\newline \newline Keywords: A. Superconductors; C. Crystal structure
and symmetry; D. Electronic band structure; D. Electron correlations
\end{abstract}

The unprecedented discovery of superconductivity in the simple binary material
MgB$_{2}$ \cite{1} has opened up new avenues to attack the problem of
maximizing \ $T_{c}$ and the symmetry of the order parameter. In the present
paper we present a model to explain the superconducting state in MgB$_{2}$
which is based on the correlated $\sigma$- and the non-correlated $\pi$-band.
The negatively charged boron layers and the positively charged magnesium
layers provide a lowering of the $\pi$-band with respect to the $\sigma$-band,
as pointed out in Ref. \cite{2} and which was also noticed first in band
calculations \cite{3,3b,3c}. Tight-binding estimates show that the bonding
$\sigma$-band, close to the Fermi level, is doubly degenerate (it has
$E$-symmetry \cite{4,5}). The quasi-2D $\sigma$-electrons are more localized
than the 3D $\pi$-electrons. This leads to an enhancement of the on-site
electron correlations in the system of degenerated $\sigma$-electrons. They
are taken to be infinite, while the somewhat weaker intersite Coulomb
interactions and electron-phonon interactions simply shift the on-site
electron energies. After carrying out a fermion mapping to $X$-operators the
Hamiltonian becomes:
\begin{align}
H &  =%
{\displaystyle\sum\limits_{p,s}}
t^{(\sigma)}(p)X^{so}(p)X^{os}(p)+V%
{\displaystyle\sum\limits_{\left\langle i,j\right\rangle }}
n_{\sigma}\left(  i\right)  n_{\sigma}\left(  j\right)  -r%
{\displaystyle\sum\limits_{i}}
n_{\pi}\left(  i\right)  +\nonumber\\
&  +%
{\displaystyle\sum\limits_{p,s}}
\varepsilon_{0}^{(\pi)}(p)\left[  \pi_{s}^{+}(p)\pi_{s}(p)+H.c.\right]  -\mu%
{\displaystyle\sum\limits_{i}}
\left[  n_{\sigma}\left(  i\right)  +n_{\pi}\left(  i\right)  \right]
.\tag{1}%
\end{align}
It is written in the usual notations, where the non-diagonal $X$-operators
describe the on-site transitions of correlated $\sigma$-electrons between the
one-particle ground states (with spin projection $s=\pm$) and the empty polar
electronic states ($0$) of the boron sites and $\mu$ is the chemical
potential. The wide $\pi$-band is shifted with respect to the $\sigma$-band by
an energy $r$. This parameter comprises the mean-field $\pi$-electron
interactions, the electron-phonon interactions (e.g. the contribution of the
strong coupling of \ $E_{2g}$-phonons with the $\sigma$-electrons \cite{6})
\textit{etc}. We included in Eq. $\left(  1\right)  $\ also the nearest
neighbour Coulomb repulsion $V$\ of $\sigma$-electrons, which is essential for
the low density of hole-carriers present in MgB$_{2}$. The mutual hopping
between $\sigma$- and $\pi$-electrons is assumed to be negligeable due to the
characteristic space symmetry of the orbitals involved. A schematic view of
our model is presented in Fig. 1. The diagonal $X$-operators ($X^{\cdot}\equiv
X^{\cdot\cdot}$) conform to the completeness relation, $X^{0}+4X^{s}=1$, and
their thermodynamic averages ($\left\langle X^{0}\right\rangle $,
$\left\langle X^{s}\right\rangle $) are the Boltzmann populations of the
energy levels of the unperturbed on-site Hamiltonian in $\left(  1\right)  $.
Due to the orbital and spin degeneracy the $\sigma$-electrons occupy their
one-particle ground state with density $n_{\sigma}=4\left\langle
X^{s}\right\rangle $ per boron site. The correlation factor for the
degenerated $\sigma$-electrons is
\begin{equation}
f=\left\langle X^{0}+X^{s}\right\rangle =1-3\left\langle X^{s}\right\rangle
=1-\frac{3}{4}n_{\sigma}\text{,}\tag{2}%
\end{equation}
and it plays an important role in all arithmetics starting from their
unperturbed (zeroth order) Green's function, $D_{\sigma}^{\left(  0\right)
}\left(  \omega\right)  =f/\left(  -i\omega_{n}-\mu\right)  $, whereas for
$\pi$-electrons $D_{\pi}^{\left(  0\right)  }\left(  \omega\right)  =1/\left(
-i\omega_{n}-\mu-r\right)  $. Note that for the conventional Hubbard model the
correlation factor in the paramagnetic phase is $1-n/2$ (see \cite{7} and
Refs. therein). The energy dispersion of both bands, $\xi\left(  p\right)
=ft\left(  p\right)  -\mu$ and $\varepsilon\left(  p\right)  =\varepsilon
_{0}\left(  p\right)  -r-\mu$, are governed by zeros of the inverse Green's
function
\begin{align*}
D^{-1}\left(  \omega,p\right)   &  =\text{diag}\left\{  D_{\sigma}^{-1}\left(
\omega,p\right)  ;D_{\pi}^{-1}\left(  \omega,p\right)  \right\}  =\\
&  \text{diag}\left\{  \frac{-i\omega_{n}+\xi\left(  p\right)  }{f}%
;-i\omega_{n}+\varepsilon\left(  p\right)  \right\}  ,
\end{align*}
which follows from the Dyson equation
\[
D^{-1}\left(  \omega,p\right)  =D^{\left(  0\right)  -1}\left(  \omega\right)
+\widehat{t}\left(  p\right)  \text{,}%
\]
to first order with respect to the tunneling matrix $\widehat{t}\left(
p\right)  =$diag$\left\{  t\left(  p\right)  ;\varepsilon_{0}\left(  p\right)
\right\}  $, where diag$\left\{  A;B\right\}  =\left(
\begin{array}
[c]{cc}%
A & 0\\
0 & B
\end{array}
\right)  $.

The chemical potential $\mu$\ for the MgB$_{2}=$Mg$^{++}$B$^{-}\left(
p^{2}\right)  _{2}$=Mg$^{++}$B$^{-}\left(  p^{n_{\sigma}}p^{n_{\pi}}\right)
_{2}$-system (Eq. $\left(  1\right)  $) obeys the equation for the total
electron density per boron site
\begin{equation}
n_{\sigma}+n_{\pi}=2,\tag{3}%
\end{equation}
with the partial electron densities given by
\begin{equation}
n_{\sigma,\pi}=2T%
{\displaystyle\sum\limits_{n,p}}
e^{i\omega\delta}D_{\sigma,\pi}\left(  \omega,p\right)  \equiv n_{\sigma,\pi
}\left(  r,\mu\right)  .\tag{4}%
\end{equation}
The electron densities, Eq. $\left(  4\right)  $, comply with the requirements
$0<n_{\sigma}<1$ (due to the correlation factor, Eq. $\left(  2\right)  $,
providing the quarter-fold narrowing of the degenerate $\sigma$-band) and
$0<n_{\pi}<2$. For any energy difference $r$\ between the\ $\sigma$- and
\ $\pi$-bands the chemical potential $\mu$\ has to be such that Eq. $\left(
3\right)  $ for the band fillings is satisfied.

From the system of Eqs. $\left(  3\text{, }4\right)  $ for a flat density of
electronic states (DOS hereafter) $\rho_{\sigma,\pi}\left(  \varepsilon
\right)  =\left(  1/2w_{1,2}\right)  \theta\left(  w_{1,2}^{2}-\varepsilon
^{2}\right)  $ with half-bandwidths $w_{1}$ and $w_{2}$ for $\sigma$- and
$\pi$-electrons, respectively and $\theta(x)$ is the theta function) we derive
the chemical potential of the $A^{2+}B\left(  p^{n_{\sigma}}p^{n_{\pi}%
}\right)  _{2}$ systems as
\begin{equation}
\mu=\frac{w_{2}-5r}{5w_{1}+4w_{2}}w_{1.}\tag{5}%
\end{equation}
The non-correlated $\pi$-electrons play the role of a reservoir for the
$\sigma$-electrons. In the energy dispersion $\xi\left(  p\right)  $\ for the
$\sigma$-electrons, the correlation factor, Eq. $\left(  2\right)  $, can be
expressed also via the electron structure parameters $w_{1},w_{2}$ and $r$ as
$f=\left(  2w_{1}+w_{2}+3r\right)  /\left(  5w_{1}+4w_{2}\right)  $.

The anomalous self-energy for the $\sigma$-electrons (Fig. 2) is written
self-consistently as
\begin{equation}
\overset{\vee}{\Sigma}\left(  p\right)  =T%
{\displaystyle\sum\limits_{n,q}}
\Gamma_{0}\left(  p,q\right)  \frac{\overset{\vee}{\Sigma}\left(  q\right)
}{\omega_{n}^{2}+\xi^{2}\left(  q\right)  +\overset{\vee}{\Sigma}\left(
q\right)  },\tag{6}%
\end{equation}
where the vertex $\Gamma_{0}$ is determined by the amplitudes of the kinematic
and Coulomb interactions such that $\Gamma_{0}\left(  p,q\right)
=-2t_{q}+V\left(  p-q\right)  $. We do not include the other kinematic
vertices in $\Gamma_{0}$ essential for a comparative analysis of the
anisotropic singlet and for triplet pairings and the energy band
renormalization at a moderate concentration of carriers \cite{7}. In momentum
space the Coulomb repulsion between the nearest neighbours (Eq. $\left(
1\right)  $) reflects the tight-binding symmetry of the boron honeycomb
lattice, $\sqrt{3+2\left(  \cos p_{y}+2\cos\frac{p_{y}}{2}\cos\frac{\sqrt
{3}p_{x}}{2}\right)  }$. Near the $\Gamma-A$ line of the Brillouin zone the
Coulomb vertex can be factorized as
\[
V\left(  p-q\right)  =2\beta t\left(  p\right)  t\left(  q\right)  ,
\]
where the parameter $\beta=V/6t^{2}$ expresses the Coulomb repulsion for the
nearest $\sigma$-electrons and the energy dispersion is $t\left(  p\right)
=3t\left(  1-\frac{p_{x}^{2}+p_{y}^{2}}{12}\right)  $. Putting the explicit
form of the vertex $\Gamma_{0}$ in Eq. (6), and after summation over the
Matsubara frequencies $\omega_{n}=\left(  2n+1\right)  \pi T$ one obtains
\begin{equation}
\overset{\vee}{\Sigma}\left(  p\right)  =%
{\displaystyle\sum\limits_{q}}
t\left(  q\right)  \left(  1-\beta t\left(  p\right)  \right)  \overset{\vee
}{\Sigma}\left(  q\right)  \frac{\tanh\sqrt{\xi^{2}\left(  q\right)
+\overset{\vee}{\Sigma^{2}}\left(  q\right)  }/2T}{\sqrt{\xi^{2}\left(
q\right)  +\overset{\vee}{\Sigma^{2}}\left(  q\right)  }}.\tag{7}%
\end{equation}
The search for a solution in the form $\overset{\vee}{\Sigma}\left(  p\right)
=\Sigma_{0}+t\left(  p\right)  \Sigma_{1}$ converts Eq. (7) for the
superconducting critical temperature\ and the gap in A$^{2+}$B$_{2}^{-}$ to
\begin{equation}
1=%
{\displaystyle\sum\limits_{p}}
t\left(  p\right)  \left(  1-\beta t\left(  p\right)  \right)  \frac
{\tanh\sqrt{\xi^{2}\left(  p\right)  +\Sigma^{2}\left(  p\right)  }/2T}%
{\sqrt{\xi^{2}\left(  p\right)  +\Sigma^{2}\left(  p\right)  }}\text{,}\tag{8}%
\end{equation}
with the gap function $\Sigma\left(  p\right)  =\left[  1-\beta t\left(
p\right)  \right]  \Sigma_{0}$.

Equalizing the gap function in Eq. $\left(  8\right)  $ to zero, one can
derive analytically $T_{c}$ in the logarithmic approximation for a flat DOS :
\begin{align}
T_{c} &  =\frac{w_{1}}{5w_{1}+4w_{2}}\sqrt{\left(  w_{1}+w_{2}-r\right)
\left(  w_{1}+4r\right)  }\exp\left(  -\frac{1}{\lambda}\right)  ,\nonumber\\
\lambda &  =\frac{\left(  5w_{1}+4w_{2}\right)  \left(  3+5\beta w_{1}\right)
}{\left(  2w_{1}+w_{2}+3r\right)  ^{3}}\left(  w_{2}-5r\right)  \left[
r-\frac{\beta w_{1}w_{2}-2w_{1}-w_{2}}{3+5\beta w_{1}}\right]  .\tag{9}%
\end{align}
Under the prefactor in the square root the restrictions for an energy shift
$r$ guarantee the assumed volume of the correlated $\sigma$-band, namely
$n_{\sigma}\geq0$ $\left(  r\leq w_{1}+w_{2}\right)  $ and $n_{\sigma}\leq1$
$\left(  r\geq-w_{1}/4=-t\right)  $. $T_{c}\left(  r\right)  $ is plotted in
Fig. 3\ for different values of the parameter $\beta$, reflecting the
suppression of superconductivity with an increase of the Coulomb repulsion.
The critical temperature $T_{c}$ has a non-monotonic dependence on the mutual
position $r$ of the $\sigma$- and $\pi$-bands. The MgB$_{2}$ case, $T_{c}=40$
K, corresponds to $r=0.085$ eV\ and a dimensionless value $\beta w_{1}=8$
V$/3w_{1}$.

The superconducting gap equation follows from Eq. $\left(  8\right)  $ taken
for T=0
\begin{equation}
1=%
{\displaystyle\int}
\rho_{\sigma}\left(  \varepsilon\right)  \varepsilon\left(  1-\beta
\varepsilon\right)  \frac{d\varepsilon}{\sqrt{\xi^{2}\left(  \varepsilon
\right)  +\Sigma_{0}^{2}\left(  1-\beta\varepsilon\right)  ^{2}}}%
\text{.}\tag{10}%
\end{equation}
It defines the anisotropic superconducting order parameter of $s^{\ast}$-wave
symmetry. In contrast to the isotropic $s$-wave gap, the superconducting gap
does not coincide with the amplitude $\Sigma_{0}$ of the gap function and it
should be defined by the minimal square root in Eq. $\left(  10\right)  $.

The near-cylindrical hole-like $\sigma$-Fermi surfaces in MgB$_{2}$ gives room
to calculate the superconducting DOS $\rho\left(  E\right)  =\sum_{p}%
\delta\left(  E-\sqrt{\varepsilon^{2}\left(  p\right)  +\Sigma^{2}\left(
p\right)  }\right)  $, where the gap function (cf\textit{.} Eq. $\left(
10\right)  $) is
\begin{equation}
\Sigma=\Sigma_{0}\left(  1-\beta t\left(  p\right)  \right)  =b\left(
1+a\cos^{2}\vartheta\right)  \text{,}\tag{11}%
\end{equation}
where $a=\beta w_{1}/\left(  12\left(  1-\beta w_{1}\right)  \right)  $,
$b=\Sigma_{0}\left(  1-\beta w_{1}\right)  $ and $\vartheta$ is the azimuthal
angle$.$ Then the superconducting DOS normalized with respect to the normal
DOS is
\begin{equation}
\frac{\rho\left(  E\right)  }{\rho_{0}\left(  E\right)  }=E\int\limits_{0}%
^{1}\frac{dz}{\sqrt{E^{2}-b^{2}\left(  1+az^{2}\right)  ^{2}}}.\tag{12}%
\end{equation}

For a Coulomb repulsion such that $\beta w_{1}<1$ the parameter satisfies
$a>0$ and the superconducting DOS becomes
\begin{align}
\frac{\rho\left(  b<E<\left(  1+a\right)  b\right)  }{\rho_{0}\left(
E\right)  } &  =\sqrt{\frac{E}{2ab}}K\left(  q\right)  ,\nonumber\\
\frac{\rho\left(  E>\left(  1+a\right)  b\right)  }{\rho_{0}\left(  E\right)
} &  =\sqrt{\frac{E}{2ab}}F\left(  \sin^{-1}\sqrt{\frac{ab}{\left(  E+\left(
1+a\right)  b\right)  q^{2}}};q\right)  \text{,}\tag{13}%
\end{align}
which is expressed in terms of the complete and incomplete elliptic integrals
$K$ and $F$, respectively with modulus $q=\sqrt{(E-b)/2E}$. The DOS\ $\left(
13\right)  $ has cusps at $E$=$\pm\left(  1+a\right)  b$. A similar result was
obtained in Ref. \cite{8} for a non-specified parameter $a>0$. In our case the
parameter $a$\ is controlled by the in-plane Coulomb repulsion $V$ as the
authors of Ref. \cite{4}\ noted. At $\beta w_{1}=0.92$ the DOS of Eq. $\left(
13\right)  $ (see Fig. 4, inset) reproduces Fig. 1(b) of Ref. \cite{8}.

But for our case of an enhanced Coulomb repulsion $\beta w_{1}>1$, we have to
take the parameter $a<0$ in the gap function $\left(  11\right)  $ and the
superconducting DOS (Fig. 4) is then given by
\begin{align}
\frac{\rho\left(  \left(  1-\left|  a\right|  \right)  b<E<b\right)  }%
{\rho_{0}\left(  E\right)  } &  =\frac{E}{\sqrt{\left(  E+b\right)  \left|
a\right|  b}}F\left(  \sin^{-1}\frac{1}{q};\sqrt{\frac{2E}{E+b}}\right)
,\nonumber\\
\frac{\rho\left(  E>b\right)  }{\rho_{0}\left(  E\right)  } &  =\sqrt{\frac
{E}{2\left|  a\right|  b}}F\left(  \sin^{-1}q;\sqrt{\frac{E+b}{2E}}\right)
.\tag{14}%
\end{align}
For our $a<0$ case the superconducting DOS\ $\left(  14\right)  $\ contains
two logarithmic divergencies at $E=\pm b$ and a gap in the energy range
$\left|  E\right|  <$ $\left(  1-\left|  a\right|  \right)  b$. The ''gap''
ratio is $1/\left(  1-\left|  a\right|  \right)  $.

Measurements on MgB$_{2}$ with scanning tunneling spectroscopy \cite{9,9b} and
with high-resolution photo-emission spectroscopy \cite{10} revealed the
presence of these two gap sizes. From the ratio 3.3 between the two gaps in
Ref. \cite{9} we can extract the parameters $\left|  a\right|  \approx2/3$ and
$\beta w_{1}\approx1.14$ (see Eq. $\left(  11\right)  $), whereas from data of
Ref. \cite{9b} one can derive $\beta w_{1}=1.21$. A value $\beta w_{1}=1.14$
can be estimated from Ref. \cite{10}. Point-contact spectroscopy \cite{10b}
shows gaps at 2.8 and 7 meV, from which we estimate a Coulomb repulsion
parameter $\beta w_{1}=1.16$. The recent study of energy gaps in
superconducting MgB$_{2}$ by specific-heat measurements revealed ''gaps'' at
2.0 meV and 7.3 meV \cite{10c} for which $\beta w_{1}=1.13$, and a gap ratio
$3-2.2$ \cite{10d}, for which $\beta w_{1}=1.14-1.15$. Measurements of the
specific heat of Mg$^{11}$B$_{2}$ also give evidence for a second energy gap
\cite{10e}.\ It is worth noticing that earlier Raman measurements \cite{11}
have revealed the presence not only of the discussed peak at $110$ cm$^{-1}$,
but also of an asymmetric peak at $65-60$ cm$^{-1}$(see Figs. 1 and 2 in Ref.
\cite{11}). These results correspond to a gap ratio $0.6-0.54$, from which we
estimate $\left|  a\right|  =0.41-0.45$ (cf\textit{.} Eq. $\left(  14\right)
$) and a Coulomb parameter $\beta w_{1}=1.25-1.22$ (Eq. $\left(  11\right)
$). Later Raman measurements \cite{11b} established pronounced peaks,
corresponding with gaps at 100 cm$^{-1}$ and 44 cm$^{-1}$. From these data one
can extract $a=-0.56$ and $\beta w_{1}=1.18$. At $\beta w_{1}=1$ we have
gapless like superconductivity. In this case the superconducting DOS is linear
with respect to the energy near the nodes of the superconducting order
parameter. Then the superconducting specific heat $C_{e}\sim T^{2}$ and the
NMR boron\ relaxation rate $\sim T^{3}$ at low temperatures\ (for BEDT-TTF
organic salts this is shown in Ref. \cite{12}$)$. From this point of view it
is interesting that the data of Ref. \cite{13} shows a $C_{e}\sim T^{2}$
behaviour and a deviation from the exponential BCS\ behaviour in $T_{1}%
^{-1}\left(  T\right)  $ of $^{11}B$ \cite{14} visualized in MgB$_{2}$.

In summary, we have analyzed the superconducting properties of the material
MgB$_{2}$ within the framework of a correlated model $\left(  1\right)  $. The
existing electron-phonon and non-phonon approaches to the superconducting
mechanism in MgB$_{2}$ can be separated in two groups: one pays attention to
the $\sigma$-electrons and the other to the $\pi$-electron subsystem. We have
taken into account both the correlated $\sigma$- and noncorrelated $\pi
$-electrons. Analysis of our results leads to the conclusion that
superconductivity occurs in the subsystem of $\sigma$-electrons with
degenerate narrow energy bands whereas the wide-band $\pi$-electrons play the
role of a reservoir. Superconductivity is driven by a non-phonon kinematic
interaction in the $\sigma$-band. A lot of evidences in favour of two
different superconducting gaps can be explained by anisotropic
superconductivity with an order parameter of $s^{\ast}$-wave symmetry, induced
by the in-plane Coulomb repulsion. For an enhanced interboron Coulomb
repulsion ($\beta w_{1}>1$) the logarithmic divergencies in the
superconducting DOS (Eq. $14$, Fig. 4) are manifested by a second gap in the
experiments. The kinematic mechanism of superconductivity for correlated
electrons was first proposed for high-T$_{c}$-cuprates (\cite{16}, see
\cite{7} and Refs. therein) as a non-phonon mechanism with correlated hopping
(\cite{17}, and Refs. therein). In Ref. \cite{18} it was shown that in the
strongly correlated limiting case the kinematic mechanism reproduces the
result of hole dressed superconductivity. In our approach the electron-phonon
coupling is hidden in the parameter $r$ . Therefore the pressure and isotope
effects can be explained by the dependence of $r\left(  \omega\right)  $\ on
the phonon modes. From the non-monotonic $T_{c}$ dependence on $r$\ it follows
that the MgB$_{2}$\ material is in the underdoped regime (around
$r\gtrsim-w_{1}/4$). For a fictitious system A$^{2+}$B$_{2}$, where the two
electrons are contributed by atom A, the superconducting critical temperature
increases with an $r$ increase. A pressure increase lowers the $\sigma$-band
with an $r$ decrease resulting in a negative pressure derivative of $T_{c}$ in
agreement with experiment \cite{20}. The bell shaped curve $T_{c}(r)$, with
the MgB$_{2}$ position in the underdoped regime, shows a possibility to reach
higher $T_{c}$'s in diboride materials such as A$^{2+}$B$_{2}^{-}$ with an
AlB$_{2}$ crystal structure. We suggest the synthesis of materials with
increased $r$-values (e.g. with ''negative chemical pressure'') and optimized
smaller interatomic B-B distances in the honeycomb plane.

\section*{Acknowledgments}

This work was supported by the Flemish Science Foundation (FWO-Vl), the
Concerted Action program (GOA) and the Inter-University Attraction Poles
(IUAP-IV) research program. V. A. I. acknowledges useful discussions with K.
Maki and S. Haas for sending their recent preprint.

\section*{Figure captions}

\begin{description}
\item [Fig. 1]A schematic view of the energy band diagram. The degenerate
$\sigma$-electrons are represented by a lower correlated band.

\item[Fig. 2] The anomalous self-energy $\overset{\vee}{\Sigma}\left(
p\right)  $ for $\sigma$-electrons. The solid line is an anomalous Green's
function (cf. Eq. $\left(  6\right)  $).

\item[Fig. 3] The non-monotonic dependence of $T_{c}\left(  r\right)  $ at
different magnitudes of the Coulomb repulsion (parameter $\beta w_{1}$). Here
$w_{2}/w_{1}=8$. MgB$_{2}$ is marked in the inset with a solid circle
positioned at $r=0.085$ eV\ and $\beta w_{1}=1.2$ for $w_{1}=1$ eV and
$w_{2}=8$ eV.

\item[Fig. 4] The superconducting\ DOS, normalized with respect to the normal
DOS,\ for the Coulomb parameter range $\beta w_{1}>1$. At $\beta w_{1}=0.92$,
the Maki result \cite{8} is reproduced (inset) (cf. Eq. $(13)$).
\end{description}
\end{document}